\documentclass[aps,prl,preprint,superscriptaddress]{revtex4-2}

\usepackage{graphicx}

\begin{document}


\title{Measurement of spin dynamics in a layered nickelate using  x-ray photon correlation spectroscopy: Evidence for intrinsic destabilization of incommensurate stripes at low temperatures}


\author{Alessandro Ricci}
\affiliation{Deutsches Elektronen-Synchrotron DESY, Notkestra{\ss}e 85, 22607 Hamburg, Germany}

\author{Nicola Poccia}
\affiliation{Institute for Metallic Materials, Leibniz Institute for Solid State and Materials Research IFW Dresden, 01069 Dresden, Germany}

\author{Gaetano Campi}
\affiliation{Institute of Crystallography, CNR, via Salaria Km 29.300, 00015, Monterotondo, Roma, Italy}

\author{Shrawan Mishra}
\affiliation{Advanced Light Source, Lawrence Berkeley National Laboratory, Berkeley, California 94720, USA}
\affiliation{School of Materials Science and Technology, Indian Institute of Technology, Banaras Hindu University, Varanasi 221005, India}

\author{Leonard M{\"u}ller}
\affiliation{Deutsches Elektronen-Synchrotron DESY, Notkestra{\ss}e 85, 22607 Hamburg, Germany}

\author{Boby Joseph}
\affiliation{Elettra Sincrotrone Trieste. Strada Statale 14 - km 163.5, AREA Science Park, I-34149 Basovizza, Trieste, Italy}

\author{Bo Shi}
\affiliation{Van der Waals-Zeeman Institute, University of Amsterdam, 1098 XH Amsterdam, The Netherlands}

\author{Alexey Zozulya}
\altaffiliation{present address: European XFEL, Holzkoppel 4, 22869 Schenefeld, Germany}
\affiliation{Deutsches Elektronen-Synchrotron DESY, Notkestra{\ss}e 85, 22607 Hamburg, Germany}

\author{Marcel Buchholz}
\affiliation{II. Physikalisches Institut, Universit{\"a}t zu K{\"o}ln, Z{\"u}lpicher Str.~77, 50937 K{\"o}ln, Germany}

\author{Christoph Trabant}
\affiliation{II. Physikalisches Institut, Universit{\"a}t zu K{\"o}ln, Z{\"u}lpicher Str.~77, 50937 K{\"o}ln, Germany}
\affiliation{Helmholtz-Zentrum Berlin f{\"u}r Materialien und Energie GmbH, Albert-Einstein-Str.~15, 12489 Berlin, Germany}

\author{James C. T. Lee}
\affiliation{Advanced Light Source, Lawrence Berkeley National Laboratory, Berkeley, California 94720, USA}
\affiliation{Department of Physics and Astronomy, Sonoma State University,
1801 East Cotati Avenue, Rohnert Park, CA 94928-3609 USA}

\author{Jens Viefhaus}
\affiliation{Deutsches Elektronen-Synchrotron DESY, Notkestra{\ss}e 85, 22607 Hamburg, Germany}
\affiliation{Helmholtz-Zentrum Berlin f{\"u}r Materialien und Energie GmbH, Albert-Einstein-Str.~15, 12489 Berlin, Germany}

\author{Jeroen B. Goedkoop}
\affiliation{Van der Waals-Zeeman Institute, University of Amsterdam, 1098 XH Amsterdam, The Netherlands}

\author{Agustinus Agung Nugroho}
\affiliation{Faculty of Mathematics and Natural Sciences Intitut Teknologi Bandung, Jl. Ganesha 10 Bandung, 40132, Indonesia}

\author{Markus Braden}
\affiliation{II. Physikalisches Institut, Universit{\"a}t zu K{\"o}ln, Z{\"u}lpicher Str.~77, 50937 K{\"o}ln, Germany}

\author{Sujoy Roy}
\affiliation{Advanced Light Source, Lawrence Berkeley National Laboratory, Berkeley, California 94720, USA}

\author{Michael Sprung}
\affiliation{Deutsches Elektronen-Synchrotron DESY, Notkestra{\ss}e 85, 22607 Hamburg, Germany}

\author{Christian Sch{\"u}{\ss}ler-Langeheine}
\email[]{christian.schuessler@helmholtz-berlin.de}
\affiliation{Helmholtz-Zentrum Berlin f{\"u}r Materialien und Energie GmbH, Albert-Einstein-Str.~15, 12489 Berlin, Germany}


\date{\today}

\begin{abstract}
We study the temporal stability of stripe-type spin order in a layered nickelate with X-ray photon correlation spectroscopy and observe fluctuations on time scales of tens of minutes over a wide temperature range. These fluctuations show an anomalous temperature dependence: they slow down at intermediate temperatures and speed up upon both heating and cooling. This behavior appears to be directly connected with spatial correlations: stripes fluctuate slowly when stripe correlation lengths are large and become faster when spatial correlations decrease. A low-temperature decay of nickelate stripe correlations, reminiscent of what occurs in cuprates due to a competition between stripes and superconductivity, hence occurs via loss of both spatial and temporal correlations. 
\end{abstract}


\maketitle


Nanoscale order of electronic degrees of freedom into periodic patterns of spatial modulation has been found in various correlated materials. A particularly prominent type is the order of charge and spins into stripes or density waves as found in the CuO$_2$ planes of hole-doped layered cuprates \cite{tranquada:1995,vojta:2009}, in the NiO$_2$ planes of hole-doped layered nickelates \cite{chen:1993,hayden:1992,sachan:1995,lee:2001,vigliante:1997,du:2000,ishizaka:2004,ghazi:2004,schuessler:2005,schlappa:2009}, as well as in other transition-metal oxides \cite{ulbrich:2012}. Stripes consist of a one-dimensional modulation of the hole density with the hole-rich sites acting as antiphase domain walls for the antiferromagnetic order on the hole-poor sites. For cuprates, the relation between stripes and superconductivity has been intensely discussed and a certain agreement has been reached that stripes or stripe-like charge-density-wave (CDW) order competes with superconductivity \cite{zaanen:1989,machida:1989,chang:2012,ghiringhelli:2012}. This notion is supported by the observation that superconductivity in hole-doped La$_2$CuO$_4$ is suppressed for doping levels near $1/8$ \cite{moodenbaugh:1988}, where stripe order is most prominent \cite{tranquada:1995,tranquada:1997,huecker:2011}. In turn, one finds stripes and CDW in various cuprates to decay at low-temperatures when the superconductivity sets in \cite{chang:2012,ghiringhelli:2012,blanco:2014,huecker:2013,croft:2014}. The picture of competition is completed by the observation that this stripe decay does not occur when the sample is exposed to a strong magnetic field suppressing superconductivity \cite{chang:2012,huecker:2013,wu:2011}.

Remarkably, a low-temperature decay of stripe order very much resembling what occurs in cuprates when superconductivity sets in can also be observed in X-ray diffraction experiments from non-superconducting layered nickelates with doping levels $< 1/3$: At low temperatures the characteristic superstructure peaks reflecting spin and charge order broaden and lose intensity indicating a loss of stripe correlations \cite{vigliante:1997,ishizaka:2004,ghazi:2004,schlappa:2009,ulbrich:2012,hatton:2002}. Since nickelates are not superconducting, a competition between stripe order and superconductivity cannot play a role; the mechanism for low-temperature decay of nickelate stripes has to be a different one. On the other hand, the mechanism at work in nickelates may also play a role in cases where superconductivity is present.

In order to learn how stripe order decays upon cooling when superconductivity as competing phase is absent, we study for a nickelate system not only spatial but also temporal correlations of spin stripes. We find slow fluctuations of the stripe order pattern over a wide temperature range. Like spatial stripe correlations, the fluctuations show an unusual temperature dependence: At intermediate temperatures, where stripe order is strongest developed, fluctuations are slow. At higher and lower temperatures fluctuations speed up such that fluctuation at 22~K are as fast as those 30~K below the onset temperature of spin stripe order. Our findings suggest that the spatial coherence of the stripe order defines an effective energy barrier of thermally activated fluctuations.

For our study, we used a single crystal of La$_{2-x}$Sr$_x$NiO$_4$ (LSNO) with  $x = 0.28$ \cite{supplement}. We studied stripe order by resonant soft x-ray diffraction with the photon energy tuned to the Ni $2p \rightarrow 3d$ ($L_3$) resonance energy, which makes the experiment directly sensitive to spatial modulation of electronic degrees of freedom \cite{schuessler:2005}. In layered nickelates the stripe-related superstructure peaks with the lowest momentum transfer occur at wave vectors $(2\epsilon,0,1)$ for charge order and $(1-\epsilon,0,0)$ for spin order (SO) where $\epsilon$ is the temperature dependent incommensurability value \cite{ulbrich:2012,yoshizawa:2000}. The notation refers to the commonly used $F4/mmm$ unit cell with $a = b = 5.43$ {\AA} and $c = 12.68$ {\AA}. We found $\epsilon$ to vary in the range between 0.292 and 0.298. Dynamics is investigated using X-ray photon correlation spectroscopy (XPCS) \cite{shpyrko:2007,konings:2011,chen:2013,gruebel:2004}  with coherent X-rays of the same photon energy. Static experiments were carried out at the P04 beam line of  PETRA III at DESY, the XPCS study was carried out at beamline 12.0.2 at the Advanced Light Source (ALS) in Berkeley. In both experiments an in-vacuum CCD was used to detect the scattered intensity \cite{supplement}. 

The stripe periodicity described by the inverse of $\epsilon$ changes with temperature \cite{vigliante:1997,ghazi:2004}. For the doping levels of the sample studied here, the increase of $\epsilon$ in particular upon heating eventually shifts the charge order peaks out of the Ewald sphere of our experiment \cite{supplement}, while the spin order peak remains reachable at all temperatures. We therefore focus on the dynamics of the spin component of stripe order. 

\begin{figure}
	\centering
	\includegraphics[width=1.0\columnwidth]{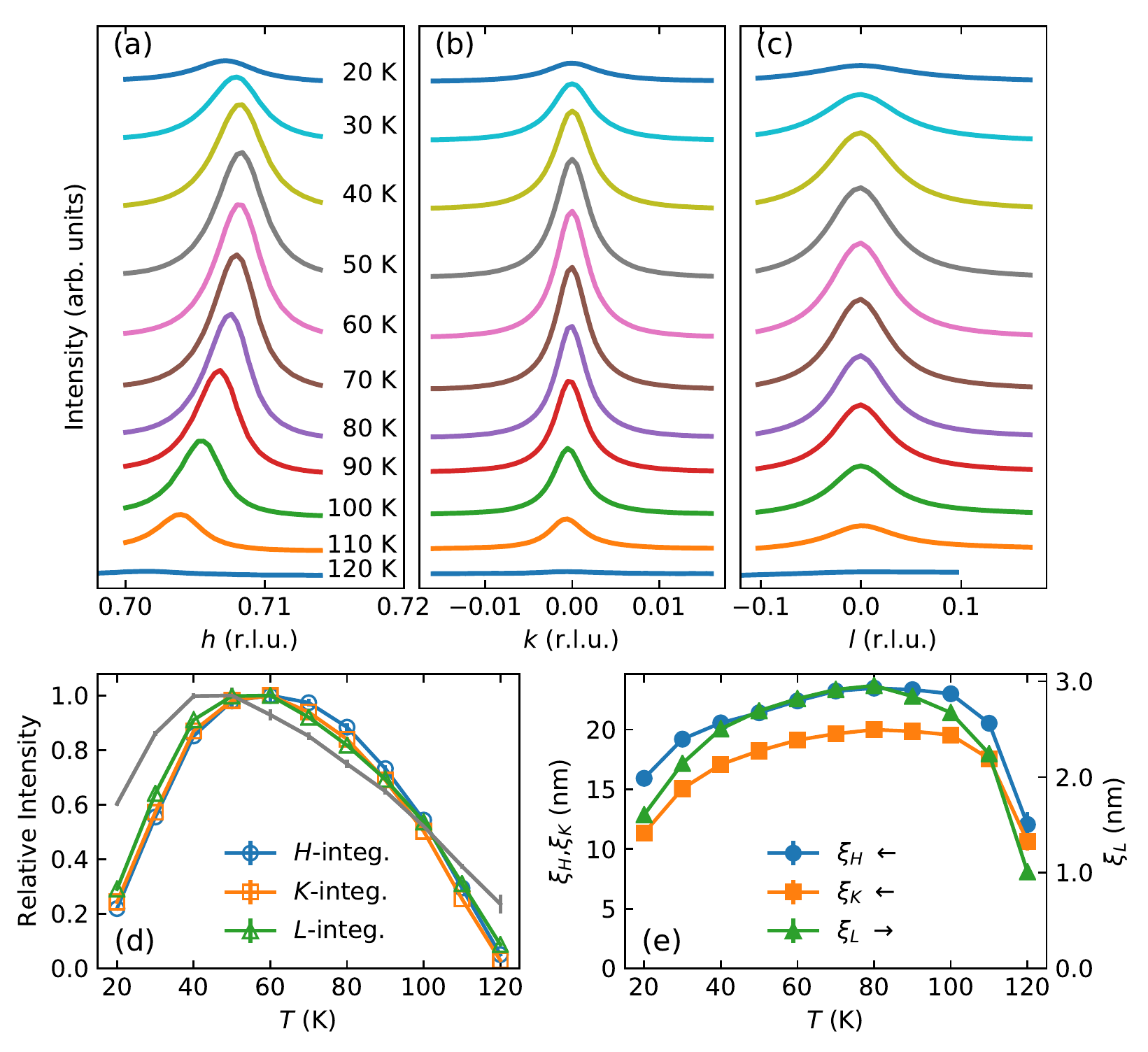}
	\caption{
		Static data from the spin stripe order. (a-c) are line cuts along different reciprocal space directions though the spin order peak at different sample temperatures. The symbols in (d) show the intensity integrated over these cuts vs. the sample temperature, the gray line in (d) is the estimated total integrated intensity. All curves in (d) have been normalized to their maximum value. The symbols in (e) show the correlation length determined from the inverse peak with. 
	}
	\label{fig:static}
\end{figure}

We start by discussing the static properties. Line cuts through the spin order peak along different directions in momentum space are presented in Fig.~\ref{fig:static}(a-c), the intensity integrated along different directions, $I_{H,K,L}$ as symbols in Fig.~\ref{fig:static}(d) and the correlation lengths extracted from the peak widths, $\xi_{H,K,L}$, in Fig.~\ref{fig:static}(e) \cite{supplement}. Spin order sets in around 120 K with a superstructure peak, which is broad in all directions. The peak growths and sharpens up to about 60 K and then starts to decay again upon further cooling. The intensity integrated along one of the reciprocal space directions at 20 K is less than a third of its 60-K value. The decrease of $I$ is accompanied by a peak broadening, i.e., a loss of spatial coherence. The broadening occurs to a similar extent in all reciprocal space directions with the correlation length along the $L$-direction being about 10 times shorter than along the other two directions, reflecting the weak correlations between the different nickel oxide planes. 

A decreasing correlation length can be either due to the formation of dislocations in the order pattern or due to a shrinking of the ordered regions into smaller and smaller patches. An indication what kind of mechanism is at work here can be obtained from the total ordered volume reflected in the integrated intensity of the scattering peak. We estimated this quantity by taking the area of the line cut along $H$ multiplied by the peak widths along $K$ and $L$. The result is presented as grey line in Fig.~\ref{fig:static}(d). While the integrated intensity varies less than the intensity of the line cuts, also the integrated quantity decays towards low temperatures. In our experiment, we are sensitive to only one of the two possible stripe orientations, namely that one where the stripe propagation vector lies in the scattering plane. A change of integrated intensity could in principle be achieved by a redistribution of the ordered volume between the two stripe directions, but since there is no obvious symmetry breaking mechanism that would favor one direction over the other, such a scenario appears unlikely. Possible scenarios that could explain the observed behavior of a loss of integrated intensity and correlation length at low temperatures are (a) the total stripe ordering is actually shrinking towards low temperatures via the formation of regions that are disordered or have a completely different kind of order or (b) defects form within the stripe-ordered patches that either reduce the ordered volume or imply a spatial phase shift, which, by destructive interference, reduces the diffracted intensity. Finally, (c), fluctuations of the stripe order may reduce the detected intensity. 

\begin{figure}
	\centering
	\includegraphics[width=0.85\columnwidth]{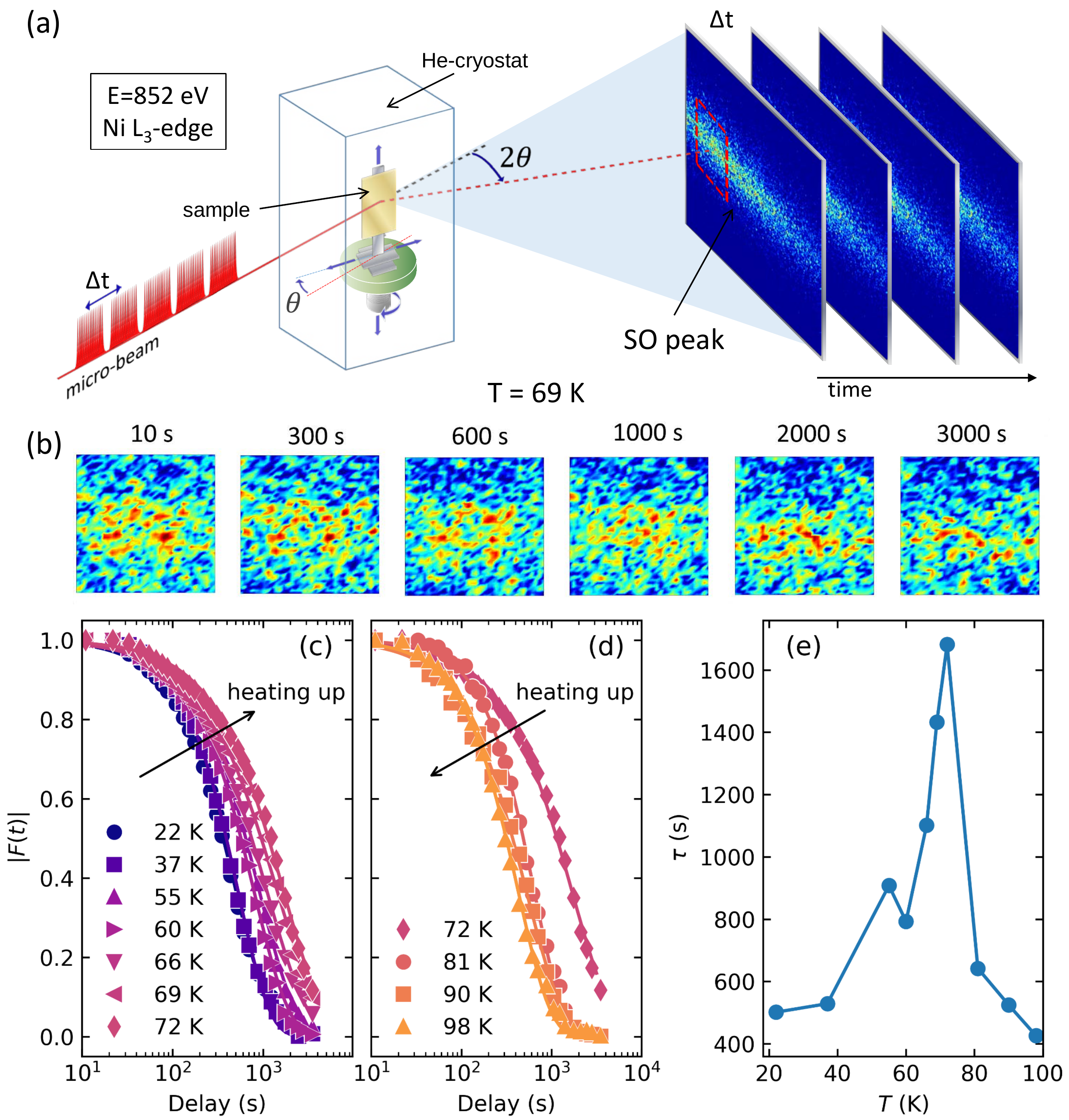}
	\caption{
		Dynamical data. (a) experimental setup for the XPCS experiment. A coherent X-ray beam hits the sample. The spin stripe diffraction peak on the detector breaks up into speckles that reflect the disorder in the sample. A series of detector images allows to analyze the fluctuations in the spatial arrangement of the stripe pattern. The region inside the red frame was used for analysis. (b) shows a series of speckle patterns recorded at 69 K near the peak maximum for different delay times. The larger the delay, the more the speckle pattern differs from the initial one. (c,d) show the autocorrelation function vs. delay for different temperatures. Symbols are experimental data, the solid lines are fits. (c) contains the lower temperatures up to 72 K and (d) the higher temperatures starting with 72 K. The characteristic fluctuation times extracted from the data in (c) and (d) are displayed in (e) vs. the sample temperature showing non-monotonous behavior.
	}
	\label{fig:dynamic}
\end{figure}

In order to address this last point, we studied the dynamics of stripe order. When illuminated with X-rays with a longitudinal and lateral coherence length that matches the illuminated sample volume, the SO peak breaks up into a myriad of speckles. The speckle pattern reflects disorder in the sample. It is caused by  interference between stripe order in different sample regions. Depending on their spatial arrangement they contribute to the diffraction signal with a different phase factor, which leads to constructive and destructive interference resulting in a characteristic speckle pattern on the detector. Any change of the spatial arrangement of these regions or their internal stripe order causes a change of the speckle pattern; its temporal evolution hence directly reveals the dynamics within the stripe order on the time scale of the measurement. An XPCS experiment uses this effect by recording a series of speckle patterns and analyzing the changes between them. The experimental scheme is depicted in Fig.~\ref{fig:dynamic}(a); a representative series of speckle patterns obtained at 69 K is shown in Fig.~\ref{fig:dynamic}(b): The later speckle patterns were recorded, the more they differ from those taken at the beginning of that series, thus indicating a temporal change of the stripe order. To quantify this evolution, we determined the intensity autocorrelation function $g_2$ which leads to the intermediate scattering function or autocorrelation function $|F(Q,t)|$ through
\begin{equation}
g_2 (t)=\frac{\langle I(\tau)I(\tau+t) \rangle_\tau}{\langle I(\tau) \rangle_{\tau}^2} = 1 + A |F(Q,t)|^2.
\end{equation}
$\langle \mathellipsis \rangle_\tau$ denotes the integration over the whole set of frames recorded for one temperature. We restrict our analysis to the central part of the SO peak with highest intensity. There we found no indications for different temporal behavior in different regions, which is why we integrate the autocorrelation function over different values of $Q$ near the peak center \cite{supplement}. We collected time-series for several temperatures; $|F(t)| := \langle |F(Q,t)|\rangle_Q$, which are presented in Fig.~\ref{fig:dynamic}(c) for $T \leq 72$ K and Fig.~\ref{fig:dynamic}(d) $T \geq 72$ K. 

All curves show a characteristic decay to zero for long delays: Pairs of speckle patterns recorded within a shorter time window are very similar (highly correlated), those with longer relative delays differ more, and those with long delays become completely uncorrelated. The time scale of this correlation loss can be inferred from the position of the steep slope in the curves. It depends on the sample temperature in a non-monotonous way with the longest decay times for 72 K. Obviously the stripe order pattern is most stable around that temperature.

In order to quantify this behavior, we describe the autocorrelation function by the exponential Kohlrausch-Williams-Watts (KWW) model: $|F(t)|=\exp(-(t / {\tau})^\gamma)$ \cite{madsen:2010}, where $\tau$ is the characteristic time-scale of the dynamics and $\gamma$ is the so-called stretching exponent. The result of least-squares-fits to the data are shown as solid lines through the data points in Fig.~\ref{fig:dynamic}(c) and (d). $\gamma$ shows no temperature dependence and scatters around 1.1 ($1.1 \pm 0.13$); the results for $\tau$ are summarized in Fig.~\ref{fig:dynamic}(e). $\tau$ shows a pronounced non-monotonous temperature dependence and changes quite strongly. As compared to its value at 100 K, the fluctuation time grows by more than a factor of four when cooling to 72 K, decreases upon further cooling and appears to approach a constant level below 30 K. 

The observation that fluctuations are slowest in a temperature range where the correlation lengths are longest suggests that correlation times and lengths are connected. Such connections are not uncommon. For Dy metal a monotonous relation between correlation length and fluctuation time has been reported \cite{chen:2013} and, generally, one can expect fluctuation times to increase when the correlation volume of the fluctuations gets larger \footnote{This is the mechanism behind critical slowing down at second-order phase transitions: Near the critical temperature, correlation lengths become large and fluctuations slow.}. 

\begin{figure}
	\centering
	\includegraphics[width=0.85\columnwidth]{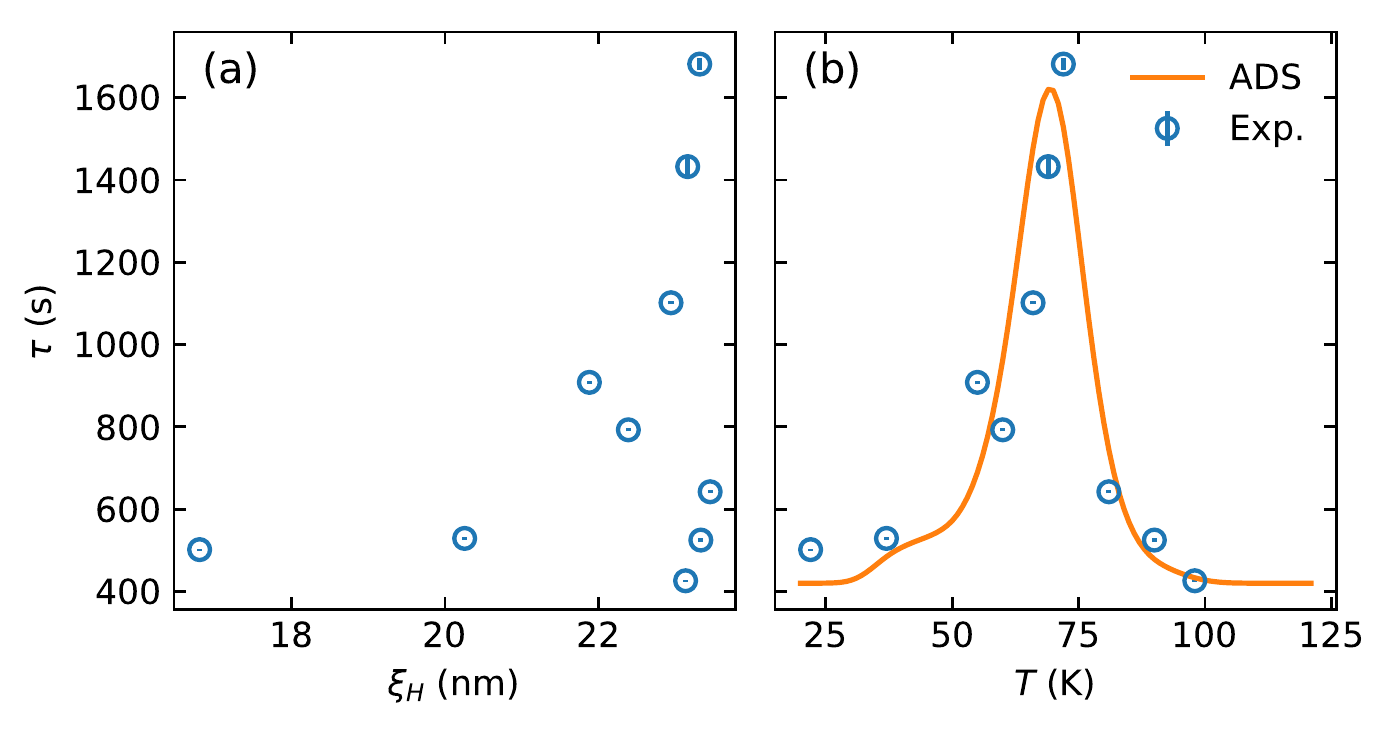}
	\caption{
		Development of the fluctuation time, $\tau$, vs. the correlation length $\xi_H$ shows no systematic behavior (a). $\tau$ vs. the sample temperature can be described by a modified activated dynamical scaling (ADS) behavior where the energy barrier for thermally activated fluctuations changes with the correlation length to the power of 2.2. 
	}
	\label{fig:ads}
\end{figure}

In the present case, however, the fluctuations are not determined by the correlation length alone. In Fig.~\ref{fig:ads}(a) we plot the fluctuation time, $\tau$, vs. the correlation length $\xi_H$ (interpolated from the results in Fig.~\ref{fig:static}(e)). The data show no clear trend; in particular fast fluctuations (small $\tau$) at low temperatures occur in the presence of shorter correlation lengths than the similarly fast fluctuations at high temperatures. This observation suggests that both temperature and correlation length define the fluctuation times. 

A model that quantitatively relates fluctuation times to both temperature and correlation length is the so-called activated dynamical scaling (ADS) \cite{karmakar:2014}. This model has been developed to describe systems where two states of almost degenerate energy form, separated by a distribution of energy barrier heights determined by randomness \cite{fisher:1987}. The ADS model predicts the fluctuation time to follow $\tau \propto \exp (C \xi ^z / T)$ where $C$ is a constant, $\xi$ the correlation length, and $z$ the so-called dynamical critical exponent, which is typically around 2. In this model $k_B C \xi^z$ acts as an effective energy barrier height for thermally activated fluctuations ($k_B$ is the Boltzmann constant). 

An ADS-like model indeed reproduces the peak in $\tau(\xi,T)$ around 72 K. The orange line in Fig.~\ref{fig:ads}(b) shows this for $\xi_H$; we obtain similar curves for $\xi_K$ and $\xi_L$. Since the bare ADS model leads to zero fluctuation times (infinitely fast dynamics) for very high and very low temperatures, we include an additive offset, $\tau_1$, to account for the finite fluctuation time that we observe for all temperatures. We further modified the ADS model by assuming an additional contribution to the correlation length, $\xi_H^0$ that has no influence on dynamics (possibly related to structural defects in the sample). The curve in Fig.~\ref{fig:ads}(b) shows the experimental data with a modified ADS curve, $\tau = \tau_0 \exp(C (\xi_H - \xi_H^0)^z /T) + \tau_1$  with $\xi_H^0 = 140$~{\AA} and $\tau_1 = 420$ s; $z$ is 2.2. Already with such simple assumptions we can model our experimental data fairly well \footnote{We expect fluctuations to speed up more upon further heating. Describing these would require further modifications of the ADS model like a temperature dependent $\tau_1$.}.  

So far we have established the existence of slow stripe fluctuations over a wide temperature range. Their time scale is at least partially determined by thermal activation over an energy barrier, which grows roughly quadratic with the correlation length. There are two types of fluctuations that match the ADS model of two (almost) degenerate states: changes of the stripe direction in a certain region or phase shifts in the stripe pattern. Direction changes could happen via motions of domain walls separating patches with different spin orientation or by whole patches changing their stripe orientation. For phase shifts within stripe patterns the overall stripe pattern and direction is conserved but the position of hole-rich and hole-poor sites would change with respect to the underlying lattice. (Such ‘phason’ modes have been observed in laser-induced stripe dynamics \cite{lee:2012}.) We expect both processes to occur, but we see no way to separate their contributions in our experiment. 

A fluctuation time scale of several tens of minutes may appear not too relevant for most material properties and cannot explain the loss of integrated stripe order scattering intensity at low temperatures, either. But our finding, while showing that fluctuations do exist, does not exclude the existence of faster fluctuations. With the coherent X-ray flux available for our experiment we could only observe speckle dynamics in the most intense part of the diffraction peak, which relates to fluctuation on long length scales. Fluctuations on shorter length scales would appear in the outer wings of the diffraction peak profile, where in our experiment the intensity was too low to determine temporal correlations. In fact, microdiffraction experiments from stripe order found a distribution of stripe domains over a wide range of length scales \cite{campi:2019}. It is hence to be expected that smaller domains and disorder on shorter length scales exist. With the observed relation between correlation length and fluctuation time, short-length correlations could be connected with much faster fluctuations \footnote{This speculation appears to contradict the $\tau_1$ in the modified ADS, but this was used to describe the fluctuations on long length scales. On shorter length scales the behavior may be different.}. 

We studied here only fluctuations of spin stripes. Since both spin and charge order show a similar low-temperature decay \cite{supplement}, one may assume that our results reflect a property of the stripe order as a whole. We note, however, that in a similar experiment for La$_{1.875}$Ba$_{0.125}$CuO$_4$ no charge stripe fluctuations could be observed \cite{chen:2016}.

As the low-temperature decay of stripe order and the concomitant speeding up of fluctuations in this nickelate cannot be caused by a competing superconducting phase, there must be another mechanism at work here. One possibility discussed in the literature is a competition between incommensurate stripe order and the periodic potential of the underlying lattice \cite{mu:2003,schmalian:2000,bogner:2001}. While ‘ideal’ stripe order favors a stripe periodicity given by the doping level, the lattice potential (and possibly the disorder potential of the randomly distributed dopant ions) may lead to a locking in of stripes to commensurate positions and a broadening of the diffraction peak \footnote{We note that, e.g., the scenario discussed in Ref. \onlinecite{mu:2003} is that of a temperature quench to low temperatures locking stripes to commensurate positions. We are in our experiment far from quenching conditions with slow temperature changes (cooling speeds are of the order of 10 K/min) and we find the peak broadening at low temperatures to be largely independent from the temperature history of the sample.}. A low-temperature change of $\epsilon$ towards commensurate $1/3$ as found here \cite{supplement} seems to accompany the low-temperature decay in most observations \cite{ishizaka:2004,ghazi:2004,schlappa:2009,ulbrich:2012,hatton:2002}, suggesting that both effects are related.

One may wonder to what extent the effect observed here might matter for the interplay between superconductivity and stripes or density wave order in cuprates. In a delicate energy balance between superconductivity and stripe/density wave order even weak additional effects might play a role. When incommensurate static stripe order already intrinsically decays into fluctuating disorder at low temperatures, this decay may help to tip the balance towards the formation of superconductivity as a competing phase. 

In summary, we observed fluctuations in the incommensurate spin-stripe pattern of La$_{1.72}$Sr$_{0.28}$NiO$_4$. These fluctuation appear to be connected with the spatial correlation of stripe order with an energy barrier for thermally activated fluctuations scaling with the approximately squared correlation length. Stripe order is spatially most correlated and temporally most stable at about half of the ordering temperature and upon further cooling becomes spatially and temporally destabilized. This process resembles the low-temperature decay of stripes or charge density waves in cuprates caused by the onset of superconductivity but here seen in the absence of such a competing phase. Low-temperature decay appears to be an intrinsic property of incommensurate stripe order.

A.R. conceived the project and designed all the experiments; C.S.-L. and M.S. contributed to the planning of the experiments; the samples were grown and characterized by A.A.N., M.Bu., C.T. and M.Br.; the static resonant x-ray diffraction measurements were performed at P04 by A.R., M.Bu., J.V. and C.S.-L.; resonant x-ray photon correlation spectroscopy experiments have been performed at ALS by A.R., N.P., S.M., L.M., B. S., J.L. and S.R.; data analysis has been done by A.R. with the contribution of M.S. and C.S.-L.; A.R., M.Br., M.S. and C.S.-L. discussed the results and worked on the interpretation of the data; A.R. and C.S.-L. have written the manuscript with the help of N.P., G.C. and M.Br. and contributions from all authors.

\begin{acknowledgments}
Project was supported by the Helmholtz Virtual Institute Dynamic Pathways in Multidimensional Landscapes. We acknowledge ALS and DESY synchrotron facilities for allocation of beamtimes and support. We thank beamline scientists for experimental help. N.P. acknowledges the Italian Ministry for Education and Research and Marie Curie IEF project for partial financial support. A.A.N. acknowledges funding from Ministry of Research, Technology and Higher Education through Hibah WCU-ITB. This work was supported by the Deutsche Forschungsgemeinschaft (DFG, German Research Foundation)—Project No. 277146847—CRC 1238 (project A02 and B04) and by the German Ministry for Science and Education through contract 05K10PK2. Moreover, support by the DFG within SFB 925 is gratefully acknowledged.
\end{acknowledgments}


%

\end{document}